\documentstyle[aps,prl,twocolumn,epsfig]{revtex}

\def\kb{k_{\rm B}}
\begin{document}
                                                          
\title{Fundamental Time Constant for a Biased Quantum Dot in the Kondo
Regime} 
 \author{Martin Plihal and David C. Langreth}
\address{Center for Materials Theory,
Department of Physics and Astronomy,
Rutgers University, Piscataway, New Jersey 08854-8019}
\author{Peter Nordlander}
\address{
Department of Physics and Rice Quantum Institute,
Rice University, Houston, Texas 77251-1892 }

\maketitle
\thispagestyle{empty}

\begin{abstract}
It is shown that the rate limiting time constant for the formation
of the Kondo state in a quantum dot can be 
extracted analytically from known perturbation theoretic results.
The prediction obtained is verified via numerical simulations
in the noncrossing approximation.
\end{abstract}


\narrowtext

{
 The basic time scale for equilibration processes in
a Kondo system has eluded researchers for more than thirty five
years.  
The Kondo effect has recently been unambiguously observed
in quantum dots  
\cite{GoldhaberetAl98Nature,CronenwettetAl98Science}.
These are prototype Kondo systems, which are also important in their
own right, and appropriate time dependent experiments
should soon appear.
The observation of Kondo resonance in the
tunneling from an STM tip via a surface adsorbate atom
\cite{madhaven,li}   has further increased the excitement.
Impurity models incorporating Kondo physics form
 one of the basic building
blocks for theories of highly correlated systems. 
 The recent success in
incorporating such theories \cite{savrasov}
 into the basic apparatus
used in electronic-structure calculations, means that
exciting new progress in this area is close at hand.
Here, we provide a surprisingly simple and elegant
solution to the time-scale problem.
}

A half a century ago, the framework and starting point for
discussion of the time scale $\tau$ for a localized spin in
an electron sea was set by Korringa \cite{korringa},
whose contribution has been promulgated at the textbook
level for decades \cite{slichter}.
  In simple terms  $1/\tau$ gives the
fractional rate at which a component of 
spin ${\bf S}$ representing a magnetic impurity 
(or a quantum dot) is  changing due to the
electrons in the conduction band (or in the leads of a quantum dot).
It is given by an expression of the type
\begin{equation}
\frac{\hbar}{\tau}= \alpha\, \kb T,
\label{eq:korringa}
\end{equation}
where $ \alpha $  is a dimensionless constant, $\kb$ is Boltzmann's
constant,
and $T$ is the temperature. 

The simplest
model that contains the relevant physics is the
Kondo model Hamiltonian
$ %\begin{equation}
 \sum_{i=1}^n( \epsilon_{k_i} - {J}{\bf{s}}_i\cdot{\bf S}),
$ %\end{equation}
where $k_i$ is a generalized  quantum number for the 
$i^{\rm th}$ conduction (or lead) electron, $ \epsilon_{k_i}$ is its
energy, and $ {\bf{s}}_i$ is its spin operator.  For a quantum dot,
$k$ is assumed to include the information on which
lead is referred to.  The exchange coupling $J$ is taken to
be independent of $k$, aside from the usual high-energy cutoff
at energies  further than $D$ above or below the Fermi level.
The Kondo temperature  $ T_{\rm K}$ is the single parameter
characterizing the physics, and  
is of order $(D/\kb ) \exp(-1/J\rho)$, where the density of
states $\rho$ is assumed constant.

The factor of $\kb T$ in Eq.~(\ref{eq:korringa}) arises because a change
in $\bf S$ involves the  absorption of a spin one
electron-hole pair of zero energy.  The phase space for such
an object is doubly restricted by the Pauli principle, and
produces the factor
$ %\begin{equation}
\int \! d\epsilon f(\epsilon)(1-f(\epsilon))
= \kb T
$ %\end{equation}
where $f(\epsilon)$ is the Fermi function.  Although Korringa's
original derivation applied to a nuclear spin where the interaction
with electrons is dipolar, it has been widely applied to impurity
electron spins as well, using the Kondo model for the interaction.  
To lowest order in $J$, the quantity $\alpha$ in Eq.~(\ref{eq:korringa})
is given by \cite{LangrethWilkins}
\begin{equation}
\alpha=\pi (J\rho)^2.
\label{alphakor}
\end{equation}
One should note that (\ref{alphakor}) represents a difference between
rates and cannot be calculated directly by a second-order perturbation
calculation.  It rather involves the solution of a master equation,
or, in many-body terminology, a sum of vertex corrections.  For 
$S=\frac{1}{2}$, as would be appropriate for a quantum dot,
the vertex corrections change the result only by a factor 
of order unity. 

The application of these ideas to a biased quantum dot allows
the phase space factor to be tuned in a continuous fashion
by varying the bias $V$ across the two leads of the dot,
which for simplicity we assume to be symmetric under lead interchange.
The phase space restriction factor in this case is
given by 
$\frac{1}{4}\sum_{l,l'}\int\! d\epsilon\; f_l(\epsilon)(1-f_{l'}(\epsilon))$,
where the indices $l$ and $l'$ designate which lead is referred to.
For example, the Fermi functions $f_1(\epsilon)$  and $f_2(\epsilon)$
have Fermi levels displaced by 
 $\pm e{V}/{2}$, respectively,
where $e$ is the magnitude of the electronic charge and $V$ is the bias.
The integral above can be evaluated analytically, with the result
that  Eq.~(\ref{eq:korringa}) should be replaced by
\begin{equation}
\frac{\hbar}{\tau}= \alpha\, \kb T \, F\left(\frac{eV}{\kb T}\right),
\label{eq:korringa2}
\end{equation}
where
\begin{equation}
F(x)= \frac{1}{4}\left(1+1 + \frac{x}{1-e^{-x}}
	+\frac{xe^{-x}}{1-e^{-x}} \right).
\label{eq:F}
\end{equation}
In writing Eq.~(\ref{eq:F}) we have sacrificed conciseness
to facilitate clarification of the origin of the terms in parentheses,
in terms of the wave function of the annihilated particle-hole pair.
The first two terms arise when the two components of this object
are in the same lead; in this case the existence of $V$ has no
effect on the result \cite{ColemanHooleyParcollet};
 the phase space for these processes
is still constricted,
and the contribution to $1/\tau$ is  still small.  For the third
term, the particle is on lead 1 and the hole on lead 2; here
the phase space is opened wide by $V$.  Finally for the
fourth term, the particle is on lead 2 and the hole on lead 1;
the phase space is, aside from an exponentially small tail,
closed off entirely as $V$ is increased.  So the essential
physical feature deriving from Eq.~(\ref{eq:F}) is that
the factor of $\kb T$ in Eq.~(\ref{eq:korringa}) is replaced at
large $V$ by $\frac{1}{4}eV$. The notion of an expanded phase
space is implicit in the Anderson model calculation of $\tau$ by
 Wingreen and Meir \cite{WingreenMeir94PRB} and in
a different context in the work of
 Kaminski {\it et al.} \cite{KaminskyetAl99PRL}
Here we have expanded the notion
and expressed it more succinctly, because it is one of the cornerstones
of what follows.

We now turn to the calculation of $\alpha$ 
when there is  a significant Kondo effect. The quantum dot
model will also provide a framework for a precise definition
of $\tau$.  A primary feature of the Kondo effect  is the appearance
of a resonance in the the imaginary part of the transition
or $t$ matrix, which is centered at the Fermi level, and which
extends logarithmically over many energy scales, the smallest of
which is characterized by the Kondo energy $\kb T_{\rm K}$.
The transition matrix then determines the 
 conductance $G\equiv I/V$, where here $I$ is the steady-state
current induced by a constant bias voltage $V$. 
  Thus a current measurement through a biased
quantum dot provides a direct measurement of the changes
in the Kondo resonance induced by a time dependent $J$ in the
Kondo Hamiltonian.   Specifically, we imagine that $J$, initially zero,
 is suddenly switched  to its final value.
The current $I(t)$ will increase, rapidly at first, but for large times $t$
should approach its final value of $GV$ exponentially with a constant 
rate.  
The exponent for this final
decay
should be independent of the method of excitation, and hence the
true characteristic  formation rate of the Kondo resonance.

Although the physics implicit in Eq.~(\ref{alphakor})
and the order of magnitude it gives,
has been
used 
by several authors  \cite{WingreenMeir94PRB,KaminskyetAl99PRL}
to obtain time scales of the quantum dot,
it is undoubtedly incorrect  to  expect that this expression,
valid to order $J^2$ only, can capture the time scale for a phenomenon
like the Kondo resonance, whose defining terms first appear in order
$J^3$.  Fortunately one can use previous analytic results
\cite{NordlanderetAl99PRL} to extract this fundamental time.
What was found there, was that the current response to a stepped
turning on of $J$ would be the same as  the  {\it equilibrium}
response to a time-dependent {\it effective temperature } $T_{\rm e}$
given by $ T_{\rm e}=T \coth \pi T t/2\hbar$. We start with the
tautology 
$
\delta I/I = [d(\ln G)/d(\ln T_{\rm e})]\; \delta T_{\rm e}/ T_{\rm e}.
$
Defining the fraction $f$ as the finite difference 
\begin{equation}
f \equiv  \frac{I(\infty)-I(t)}{I(\infty)},
\label{f}
\end{equation}
and
 \begin{equation}
f_0\equiv-2\,\frac{d(\ln G)}{d(\ln T)},
\label{f0}
\end{equation}
we have  approximately
$
 f  =  f_0(\coth \pi Tt/2\hbar - 1 )/2.
$
As $t$ becomes large, this approximation becomes exact, and we
obtain
\begin{equation}
 f \rightarrow  
	f_0  \exp \left(-\frac{t}{\tau}\right),
\label{dell}
\end{equation}
where the value of $\alpha$ defining $\tau$ via 
Eq.~(\ref{eq:korringa}) is given by
\begin{equation}
\alpha=\pi,
\label{alpha}
\end{equation}
instead of by Eq.~(\ref{alphakor}).  When the phase space expansion
factor (\ref{eq:F}) for finite $V$ is restored \cite{vdep},
 then the predicted $\tau$
should be given by Eq.~(\ref{eq:korringa2}).

Our result for $\tau$, although analytically derived only in perturbation
theory, contains no non-universal parameters like $J$
or $D$, and so might be expected be valid in parameter ranges 
exceeding those implied in its derivation.  However, the expressions
do not contain $T_{\rm K}$,
so corrections may appear when $\kb T$ and $eV$ become much smaller 
than $\kb T_{\rm K}$.
 For example, one could suspect
that a particle-hole pair would require an energy of
$\sim$$\kb T_{\rm K}$ in order to excite the singlet ground Kondo state,
so that $1/\tau$ would vanish exponentially as $\sim$$\exp(-T_{\rm K}/T)$
for $eV \ll \kb T \ll \kb T_{\rm K}$ instead of linearly with $T$.

\vspace*{1em}
\begin{figure}
\centerline{\epsfxsize=3.2in%0.99\columnwidth
\epsfbox{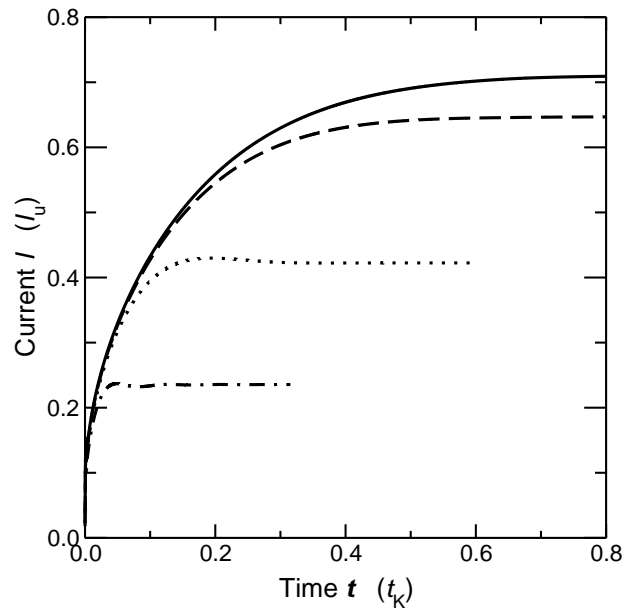}
}
\vspace{1em}
\caption{
 Instantaneous current $I$    
through the quantum dot  
versus the time $ t$ after $J$ was turned   on.
The units $I_{\rm u}=(e^2/\pi \hbar)V$ and  
$t_{\rm K} = \hbar/\kb T_{\rm K}$.
The solid, dashed, dotted, and dash-dotted curves correspond to
the respective $eV/\kb T_{\rm K}$  values of 1, 5, 20, and 80.
The temperature $T=1.5 T_{\rm K}$.
}
\end{figure}

	We have made numerical simulations\cite{method} to test the response
of the current to sudden increases in $J$ at various biases
and temperatures.   The essential feature is that
$J$ is suddenly increased from a negligible value to a
substantial value, in this case given by $J\rho=1/2\pi$.   We
found results for two temperatures: $T=1.5T_{\rm K}$ and
$T=15T_{\rm K}$, where $ T_{\rm K}$ corresponds to the 
final value of $J$ above.  A sampling of the
 data is  shown in Fig.~1.

One sees clearly that the current versus time curves have
the general features discussed above, with the time required
for saturation decreasing with increasing bias $V$.  For
$eV$ somewhat greater than   $\kb T_{\rm K}$,
small, damped oscillations
occur  \cite{PlihaletAl00PRB,ColemanUnstable,SchillerHershfield00PRB}
at the non-trivial frequency given by
$\hbar\omega=eV$.
They are caused by
transitions  between the split Kondo peaks (SKP),
which occur\cite{WingreenMeir94PRB,Sivan} for $eV$ greater than
several times $\kb T_{\rm K}$.
Such SKP oscillations
should not occur at all for the completely symmetric
mechanism  of suddenly turning on $J$.  However,
as described below, in order to make our simulations
numerically feasible, it was necessary to include
a small (asymmetric) mixed valent admixture, which is responsible
for the excitation of the SKP oscillations in this case.

\vspace{1.5em}
\begin{figure}
 \centerline{\epsfxsize=3.2in%0.99\columnwidth
\epsfbox{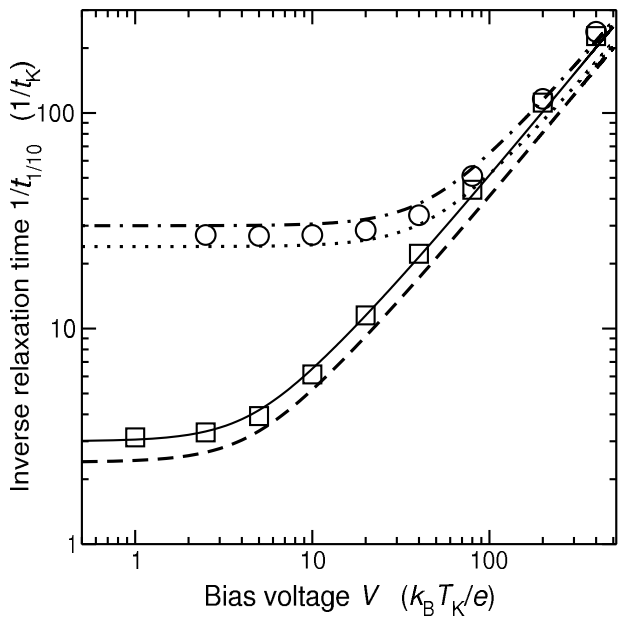}
}
\vspace{1em}
\caption
{ Comparison 
of Eq.~(\ref{eq:korringa2}) with the
simulation results. The ordinate shows the
rate $1/t_{1/10}$ as defined in the text
in units of $1/t_{\rm K}=\kb T_{\rm K}/\hbar$. For $T=1.5T_{\rm K}$,
the dashed and solid lines show the theoretical curves for  $\alpha_{1/10}$
values of 1.6 and 2.0 respectively, while the squares show the simulation
results. The dotted and dash-dotted lines and
circles respectively show the corresponding results for $T=15T_{\rm K}$.
}
\end{figure}

The scheme we adopted for analyzing the data of
\mbox{Fig.~1}, comprised first of determining the time, which we
call $t_f$, corresponding to the current fraction $f$  defined earlier.
We choose this fraction to be 1/10, an amplitude that
is sufficiently large that the SKP oscillations
do not spoil the analysis, and sufficiently small that the
ultimate time scale has been reached approximately.
One finds, by solving Eq.~(\ref{dell})  for $t$, that $t_f$ is
given  simply by replacing $\tau$ by           $t_f$
in Eq.~(\ref{eq:korringa2}), provided that $\alpha$ 
is replaced by 
$
\alpha_f \equiv  \alpha/\ln(f_0/f).
$
The amplitude $f_0$ was calculated directly both from the
large-time asymptotes of the simulations and from
Eq.~(\ref{f0}), with similar results; it was found to be
remarkably slowly varying with temperature in
the temperature range of the simulations. \cite{largeT}  Assuming that
$\alpha=\pi$, we found taking $f=1/10$ that $\alpha_f=1.6$ for both
$T=1.5 T_{\rm K}$ and $T=15 T_{\rm K}$.  In Fig.~2 we show
our simulation results along with the theoretical prediction
from Eq.~(\ref{eq:korringa2}) with $\alpha_{1/10}=1.6$.   Better
fits to the simulation can be obtained by taking
$\alpha_{1/10}= 1.8  $ (not shown) or  $\alpha_{1/10}= 2.0$;
these correspond to $\alpha$ values of $1.1\pi$ and $1.2\pi$
respectively. Such a 10--20\% difference from the prediction of
Eq.~(\ref{alpha}) is consistent
with the expected accuracy of the simulation method,
exacerbated
by the fact that the simulations contained a small
mixed valent contribution  not
present in the Kondo Hamiltonian.
The perturbative value
of $\alpha=0.03\pi$ predicted by Eq.~(\ref{alphakor}) is definitely
ruled out. The substantial agreement
gives added support to our time scale predictions
of Eqs.~(\ref{eq:korringa2}) through % (\ref{eq:F}) 
(\ref{alpha}).  The challenge
of extending our results to the  range $\kb T,eV \ll \kb T_{\rm K}$
still remains.

 We thank P. Coleman, C. Hooley,
and O. Parcollet for useful discussions.
 The work was supported in part by NSF grants DMR 97-08499 (M.P. \& D.C.L.)
and DMR 00-93079 (D.C.L.),
DOE grant DE-FG02-99ER45970 (D.C.L.), a NIST cooperative research
 grant (M.P.),
and the Robert A. Welch Foundation (P.N.).

\vspace*{\baselineskip}
{\em Note added on 27 September 2001. }
Since this work was originally posted, we became aware
of work by A. Rosch, J. Kroha, and P. W\"olfle, Phys.\ Rev.\ Lett.\ 
{\bf 87}, 156\,802 (2001).
These authors calculate the imaginary 
part of the pseudo-fermion  self-energy  in time-independent NCA.
If this quantity is closely related   
to the time-dependent  definition, Eq.~(7), of the Kondo state's
formation rate, then their work would include corrections that go beyond
the simple  argument that led to Eq.~(3),
but which would automatically be included in our NCA simulations.
It is possible that  the small deviation of the squares in Fig.~2
above the solid line for the highest three points could be due
to this. However, the accuracy achievable by  our data analysis
is limited in this region
and drawing any physical conclusion from this deviation
might be straining these limits.
Because the relative effect of the mixed valent contribution 
increases with increasing $V$, we are unable approach the
asymptotic limit exemplified by Eq.~(4) of  the 
work of Rosch et al.  
We thank Achim Rosch for correspondence.


\begin{thebibliography}{10}
\bibitem{vdep}
The implicit assumption that $\alpha$ itself is insensitive
to $V$ is consistent with its $T$ independence, but should
be a subject of future investigation.
\bibitem{method}
{The simulations used the noncrossing approximation,
a reliable calculational scheme, whose time-dependent
version we have adequately described
and referenced  previously in
  Refs.~\onlinecite{NordlanderetAl99PRL,PlihaletAl00PRB}.
It is valid in the temperature range of our simulations, but breaks
down for $T\ll T_{\rm K}$.  In the same notation as we used
previously in Ref.~\onlinecite{PlihaletAl00PRB}, 
we took Anderson model parameters
$\epsilon_{\rm dot}=-2 \Gamma_{\rm dot}$, with a parabolic band of width
$9\Gamma_{\rm dot}$.  These give an equivalent Kondo model $J\rho=1/2\pi$
and $\kb T_{\rm K}=10^{-3} \Gamma_{\rm dot}$.  The equivalence between
the models is good, but not perfect, because for 
computational-resource reasons, the small mixed valent overlap could
not be reduced to zero by choosing a much larger $|\epsilon_{\rm dot}|$.
For times $t<0$, the coupling $J$
was for practical purposes turned off by taking   
$\epsilon_{\rm dot}=-5 \Gamma_{\rm dot}$.}
\end{thebibliography}
\end{document}